# Tetris

Jiajun Xu, Sam Huang

{xujiajun, sam.huang}@yeefoundation.com

( draft 0.2 )

## Abstract

Tetris is an Asynchronous Byzantine Fault Tolerance consensus algorithm designed for next generation high-throughput permission and permissionless blockchain. The core concept of Tetris is derived from Reasoning About Knowledge, which we believe to be the most appropriate tools for revealing and analyzing the fundamental complexity of distributed systems. By analyzing the states of knowledge that each participant attained in an unreliable system, we can capture some of the basis underlying structure of the system, then help us designing effective & efficient protocols. Plus the adoption of Full Information Protocol (FIP) with the optimized message traffic model, Tetris has finally got high performance, with proved safety. Tetris achieve consensus finality in seconds, means transactions can be confirmed greatly faster than other scheme like Pow/Dpos. Tetris also achieve fairness, which is critically important in some areas such as stock market etc.

# 1. Introduction

Blockchain technology has attracted great attention as one of the biggest discoveries in the past decade for its capabilities of create trust among unlimited number of trustless nodes with no identity authentication. It has emerged as a potentially disruptive technology with great opportunity to transform our society and therefore be entitled blockchain revolution[23], nearly as important an innovation as the creation of the Internet.

However, the gap between vision and reality is still huge. Especially at the most important side of technical, current blockchains still have great room for improvement in order to achieve industrial-scale application.

For example, the most successful project of blockchain cryptocurrency, Bitcoin has long faced complaints for its hours long time waiting for confirmation of transactions. And Ethereum, a cute cat Crypto-Kitties once almost disrupted the whole network - creating network congestion and traffic for all users of the Ethereum blockchain.

The performance of blockchain significantly relies on the performance of the adopted consensus mechanisms, e.g., in terms of speed of transaction process, speed of consensus finality, robustness to arbitrary behaving nodes and network scalability.

After Pow, numerous PoX have been proposed but with few successfully go beyond the high energy consumptive Pow.

Almost all PoX, and some relative new promising technics like VRF, sound good at first, finally turn to the direction of add a BFT scheme to guarantee its safety, for example: Pos+BFT in TON, Dpos+BFT in EOS, VRF+BFT in Algorand... This kind of taking BFT as a patch to rescue the unreliable scheme comes along with potentially many problems.

Other proved schemes like PBFT, Hashgraph, capable of supporting much higher performance, but determined by its system model and assumptions, can only apply to permission blockchain.

We here take a different approach.

At first, we are trying to design a full Byzantine Fault Tolerance algorithm with proved safety at the core with the architecture adaptive to permissionless blockchain context.

BFT algorithm require a set of designated members to communicate with each other and take collaborative actions. Then, so we'll depend on an upper level protocol to choose members, rotate members, evaluate them and combine incentive into the system. We can take this protocol as pluggable. Here traditional PoX can play a role while the most important safety property was guaranteed by the BFT core.

Create a BFT algorithm is not an easy task, especially in a permissionless blockchain environment. We try to ensure that our design decision can deduced from the core requirement of the system, from the fundamental properties of the system model and environment. We found that the theory of knowledge of distributed system can play an important role in this field and reasoning about knowledge help us capture the most essential concept of system, protocol and collaboration.

When construct the protocol and algorithm, we try at every step, not only we propose how/what to do, more important, we should point out why to do? what's the benefit & shortcoming? For example, we adopt full-information protocol(FIP) in Tetris, contrary to traditional belief that FIP cause unrealistic traffic load, we achieve the minimum bandwidth requirement.

At the end, correctness proof will provided for all the key steps. It's crucial in a byzantine environment.

Although the consensus mechanism is important for the performance of blockchain, it is not all. Imagine that if you design a consensus algorithm support tens thousands of transactions per second, but every node in the system need to consume tens of megabytes per second of bandwidth and most of the cpu capacity. It is still unpractical.

So in Tetris, in order to reach the maximum potential of consensus algorithm, We have carefully designed the p2p overlay sub-network structure, push/pull traffic model and multilevel node mode etc. We will describe them later.

# 2. System Model

Blockchain at its core is a distributed system, more exactly a decentralized distributed system. When talking about distributed system, it's critical to define the system model precisely in order to create formal proofs of various guarantees of the algorithms under investigation.

We focus on asynchronous distributed system with byzantine failure model because we believe this model most accurately captures the properties of the public internet and trustless blockchain nodes on the network.

By the term *distributed*, we mean that the system is composed of a set of nodes that can communicate only by sending/receiving messages along a set of channels. The network is not necessarily completely connected.

By the term *asynchronous*, we mean that there is no global clock in the system, the relative speeds of nodes are independent, and the delivery time of messages is finite but unbounded.

The network can drop, delay, duplicate, or deliver messages out of order. However, we assume that messages are eventually delivered, provided that the corresponding senders keep on retransmitting them.

We assume *authenticated channels*, where the receiver of a message can always identify its sender. Each node has a public/private key pair and all nodes know the others' public keys. We use these keys to implement authenticated channels, and sign messages where needed.

We assume a *Byzantine failure model* where nodes may deviate arbitrarily from the protocol. We allow for a strong adversary that can coordinate faulty nodes. The adversary is assumed to be computationally bounded, meaning he cannot subvert common cryptographic techniques such as signatures. It is also assumed that secure hash functions exist, for which collisions will never be found. Typically, we assume the number of faulty nodes cannot exceed 1/3 of total node number.

The description of the system is based on the behaviors of the individual nodes/processes in the system. (The term *node* and *process* are the same meaning and either could be used later following traditional preferences in the context.) Details are here:

### *Local State*

*Local state* contain all the local information of a process.

### *Action*

An *action* is a function from *local state* to *local state*. There three types of actions: *send actions* denoted *send(m)* where m is a *message*, *receive actions* denoted *receive(m)*, and *internal actions*.

### *Event*

An *event* is a tuple $\langle s, \alpha, s' \rangle$ consisting of a state, an action, and a state. Each event corresponds to a state transition. The $j^{th}$ event in process i's history, $\langle s_i^{j-1}, \alpha_i^j, s_i^j \rangle$ is denoted $e_i^j$.

## Local History

A *local history*, $h_i$ of process i, is a (possibly infinite) sequence of alternating *local states* and *actions*, beginning with a distinguished *initial state*. We write such a sequence as follows.

$$h_i = s_i^0 \xrightarrow{\alpha_i^1} s_i^1 \xrightarrow{\alpha_i^2} s_i^2 \xrightarrow{\alpha_i^3} s_i^3 \ldots$$

If we assumed that the local state includes a description of all past actions in the local history, then the local history may be equivalently described as following:

$$h_i = s_i^0, s_i^1, s_i^2, s_i^3 \ldots$$

## Message

A *message* is a triple $\langle i, j, b \rangle$ where *i* is the sender of the message, *j* is the message recipient, and *b* is the body of the message.

## Asynchronous Runs

We identify a system with its set of possible runs, which is a complete description of all the relevant events that occur in a system over time. At every step in a run, the system is in some global state, where a global state is a description of each process's current local state.

Each asynchronous run is a vector of local histories, one per process, indexed by process identifiers. Thus we use the notation

$$a = < h_1, h_2, h_3, \ldots h_N >.$$

## Asynchronous distributed system

We model asynchronous distributed system as consists of the following sets.

1. A set $P = \{p_1, p_2, \ldots, p_N\}$ of processes/nodes, where *N* is the total number of processes in the system.

2. A set $C \subseteq \{(i, j) \mid i, j \in P\}$ of channels. The occurrence of (i, j) in *C* indicates that process i can send messages to process j.

3. A set $H_i$ of possible local histories for each process i in *P*.

4. A set *M* of messages.

5. A set *A* of asynchronous runs.

The model of an asynchronous system does not mention time. However, there is an ordering of events in the system due to the fact that certain events are known to precede other events. It is named *happens-immediately-before* and *happens-before* relation.

### *Happens-immediately-before*

Event $e_i^x$ *happens-immediately-before* event $e_j^y$, denoted $e_i^x \mapsto e_j^y$, if and only if (1) $e_i^x$ and $e_j^y$ are different events in the history of some process i and $e_i^x$ occurs earlier in the sequence, i.e. i=j and x<y, or (2) $e_i^x$ is the sending of a message and $e_j^y$ is the reception of that message; i.e. there exists m such that $e_i^x = send(m)$ and $e_j^y = receive(m)$.

### *Happens-before*

The *happens-before* relation, denoted $\rightarrow$, is the transitive closure of *happens-immediately-before*. Thus if $e_i^x \rightarrow e_j^y$, then either $e_i^x \mapsto e_j^y$ or there exists an event $e_k^z$ such that $e_i^x \rightarrow e_k^z$ and $e_k^z \mapsto e_j^y$.

### *Global State*

A global state at some time of run *a* is a N-vector of prefixes of local histories of *a*, one prefix per process.

### *Consistent Cut*

A *consistent cut* of a run is any global state such that if $e_i^x \rightarrow e_j^y$ and $e_j^y$ is in the global state, then $e_i^x$ also in the global state.

The definition of *global state* above is not always consistent, while there are chance prefix of one process's history contain reception of a message but no prefix of others history contain the sending of the message. This is meanless, so the definition of *consistent cut* in asynchronous system is analogous to the global state in synchronous system.

Note that a consistent cut is simply a vector of local states, we will use the notation *(a,c)[i]* to indicate the local state of process *i* in cut *c* of run *a*.

Most of the description above come from[5], and are conceptually similar to the timed runs model for describe synchronous system.[4][7]

## **Protocols**

Processes usually perform actions according to some *protocol*. Intuitively, a *protocol* for process *i* is a description of what actions process *i* takes as a function of its local state. A system's all possible runs must follow a protocol *P*, an *initial state*, and a *failure pattern*.

Formally, we define a protocol in terms of a **message generation function**(which describes the messages that each process sends as a function of its local state), a **state transition function**, and an **output function**.

## Full-information protocols

A protocol is said to be a full-information protocol if each process is required to send all its current state to all processes at each step. The state of a process in a full-information protocol consists of the process's name, initial state, message history, etc.

Thus the **message generation function** and **state transition function** of a full-information protocol is predetermined. So the states of processes following a full-information protocol are completely independent of their decision function; full-information protocols differ only in their **output functions**.

Intuitively, the states of the processes in a full-information protocol make the finest possible distinctions among histories. It gives each process as much information about the operating environment as any other protocol could. That is why the full-information protocol is particularly well suited for proving possibility and impossibility of achieving certain goals in distributed systems, and also for the design and analysis of distributed protocols.

Roughly, it looks as if every process send all the information of its state to all the other processes at each step can easily result in unrealistically large messages size. But in practice，we will see in section 5, the total bandwidth requirement can reduce to a minimum with the the help of hash chain and push/pull traffic mode.

# 3. Knowledge & Common Knowledge

After a clear definition of the concept of the system model and protocol, we can catch the points here:

1. The tasks that distributed systems are required to perform are normally stated in terms of the global behavior of the system. From outside of the system, we view it has a whole.

2. The actions that each individual process performs, based on the protocol, can only depend on its local information.

Therefore, the most essential problem is how can we try to obtain some level of global information or group information from the local information.

Reasoning about knowledge has been argued the right tools to analyses such systems. We regard local information of an individual process as it's knowledge. And we think of communication in the system as a means of transferring knowledge. Process receive messages to improving its state of knowledge and send messages to share it's knowledge to others. At any time, what we have is individual process's accumulating knowledge. What we want is some form of "group knowledge" derived from process's own knowledge to perform coordinative actions.

Reasoning about knowledge has often been introduced in the classical distributed system textbook by an interesting example of "the muddy children puzzle"[1][2]. And another similar "blue-eyed islander puzzle" has also recently attracted the attention of a famous mathematician, Terence Tao[25]

In [2], J.Y.Halpern defined a hierarchy of states of group knowledge, and described the relationship between *common knowledge* and a variety of desirable actions in a distributed system. The weakest state of group knowledge is *distributed knowledge*, which corresponds to knowledge that is distributed among the members of the group, without any individual process necessarily having it. The strongest state of knowledge in the hierarchy is *common knowledge*, which roughly corresponds to "public knowledge".

If process i know a given fact $\varphi$, we denote it by $K_i\varphi$. Knowledge of processes need to satisfy two properties. The first is that a process's knowledge at a given time must depend only on its local history: the information that it started out with combined with the events it has observed since then. Secondly, we require that only true things be known, or more formally:

$$K_i\varphi \supset \varphi$$

i.e., if an process i knows $\varphi$, then $\varphi$ is true.

What does it mean to say that a group G of processes knows a fact $\varphi$? here are some reasonable possibilities from [2].

$D_G\varphi$ (read "the group G has *distributed knowledge* of $\varphi$"): We say that knowledge of $\varphi$ is distributed in G if someone who knew everything that each member of G knows would know $\varphi$. For instance, if one member of G knows $\psi$ and another knows that $\psi \supset \varphi$, the group G may be said to have distributed knowledge of $\varphi$.

$S_G\varphi$ (read "*someone in G knows $\varphi$*"): We say that $S_G\varphi$ holds iff some member of G knows $\varphi$. More formally,

$$S_G\varphi \equiv \bigvee_{i \in G} K_i\varphi$$

$E_G\varphi$ (read "*everyone in G knows $\varphi$*"): We say that $E_G\varphi$ holds iff all members of G know $\varphi$. More formally,

$$E_G\varphi \equiv \bigwedge_{i \in G} K_i\varphi$$

$E_G^k\varphi$, for k $\geq$ 1 (read "$\varphi$ is $E^k$-*knowledge in* G"): $E_G^k\varphi$ is defined by

$$E_G^1\varphi = E_G\varphi$$

$$E_G^{k+1}\varphi = E_G E_G^k\varphi, \, for \, k \geq 1.$$

$\varphi$ is said to be $E^k$-knowledge in G if "everyone in G knows that everyone in /g knows that ... that everyone in G knows that $\varphi$ is true" holds, where the phrase "everyone in G knows that" appears in the sentence $k$ times.

$C_G \varphi$ (read "$\varphi$ is *common knowledge* in G"): The formula $\varphi$ is said to be common knowledge in G if $\varphi$ is $E_G^k$-knowledge for all $k \geq 1$. In other words,

$$C_G \varphi \equiv E_G \varphi \wedge E_G^2 \varphi \wedge \cdots \wedge E_G^m \varphi \wedge \ldots$$

(We omit the subscript G when the group G is understood from context.)

Clearly, the notions of group knowledge introduced above form a hierarchy, with

$$C\varphi \supset \cdots \supset E^{k+1}\varphi \supset \cdots \supset E\varphi \supset S\varphi \supset D\varphi \supset \varphi$$

The analysis of knowledge in distributed system reaches the following conclusions:[2]

1. When communication is not guaranteed it is impossible to attain common knowledge. This generalizes the impossibility of a solution to the well-known *coordinated attack* problem. It is also impossible to reach consensus in a distributed system, so in our system model in Section 2, we assume that messages should eventually delivered.

2. Common knowledge of strict definition, can not be attained in practical distributed systems, because that common knowledge can only be attained in systems that support simultaneous coordinated actions.

3. Introduced with the concept of *internal knowledge consistency*, it is safe at certain context to assume that certain facts are common knowledge, even when strictly speaking they are not. For example, in the muddy children puzzle, or a synchronous protocol that proceeds in rounds, in which it is guaranteed that no process will ever receive a message out of round.

4. There are weaker variants of common knowledge, which are attainable in practical systems. For example, $\epsilon$-common knowledge, $\Diamond$-common knowledge(eventual common knowledge), timestamped common knowledge.

# 4. Revisit Consensus

Since the first introduction of "The byzantine generals problem" by L. Lamport, R. Shostak, and M. Pease[10], Numerous articles have been published on the problem of consensus under all kinds of situations. Here we will revisit the most simple Binary Byzantine Consensus.

The consensus problem is typically defined for a set of *n* known processes, with the maximum number *t* of faulty processes. We say that a process is correct if it follows its algorithm until completion, otherwise it is said to be faulty.

A binary consensus algorithm aims to achieve consensus on a binary value v ∈ {0, 1}. Each process proposes its initial value (binary) and decides on a value v. The problem can be formally defined in terms of three properties:

- Validity: If all correct processes propose the same value v, then any correct process that decides, decides v.

- Agreement: No two correct processes decide differently.

- Termination: Every correct process eventually decides.

The first two properties are safety properties, i.e., properties that say that some 'bad thing' cannot happen, while the last is a liveness property, i.e., a property that states 'good things' that must happen.

There are many algorithms have been proposed in the past decades for this kind of Asynchronous Binary Byzantine Consensus problem. Most of them are complex, unintuitive and relatively hard to implement. So here we are trying to create such an algorithm that is mainly deduced from some basic principle and keep simple to be intuitively understood and implemented. We simplify the things by:

1. Adoption of FIP, help us mainly focus on the decision function $\mathcal{D}$.
2. Based on the maximum knowledge a single individual process can obtain, to infer the status of other processes in the system.

The decision function take the local history $h_i$ as parameter, $h_i = (a, c)[i]$ is the local history of a consistent $c$ of a run $a$. The output of the decision function in a binary protocol is either 0 or 1 or $\perp$ meaning no output currently.

$$\mathcal{D}(h_i) = \{0, 1, \perp\}$$

The decision function $\mathcal{D}(h_i)$ must satisfy:

1. With the accumulation of knowledge process attained by receive messages and growth of the history, $\mathcal{D}$ must eventually output 0 or 1. i.e., process must decide eventually.

2. After $\mathcal{D}(h_i)$ output 0 or 1, it should always output the same value later. i.e., process once decided a value, it can not change then.

3. When $\mathcal{D}(h_i)$ of process i ready to decide a value *v*, it must be determined that process i have obtain the eventual common knowledge of the fact that every processes will eventually decide on the same value. i.e., $(a, c)[i] \vDash C^{\Diamond}\varphi$ where $\varphi$ means $\mathcal{D}(h) = v$.

4. When $\mathcal{D}(h_i)$ of process i ready to decide a value *v*, it must be determined that process i know there will never has eventual common knowledge of the fact that some processes decide on the opposite value.

5. Certainly, $\mathcal{D}(h_i)$ can not always decide a fixed value of 0 or 1, which violate the validity property of consensus.

# 5. Tetris

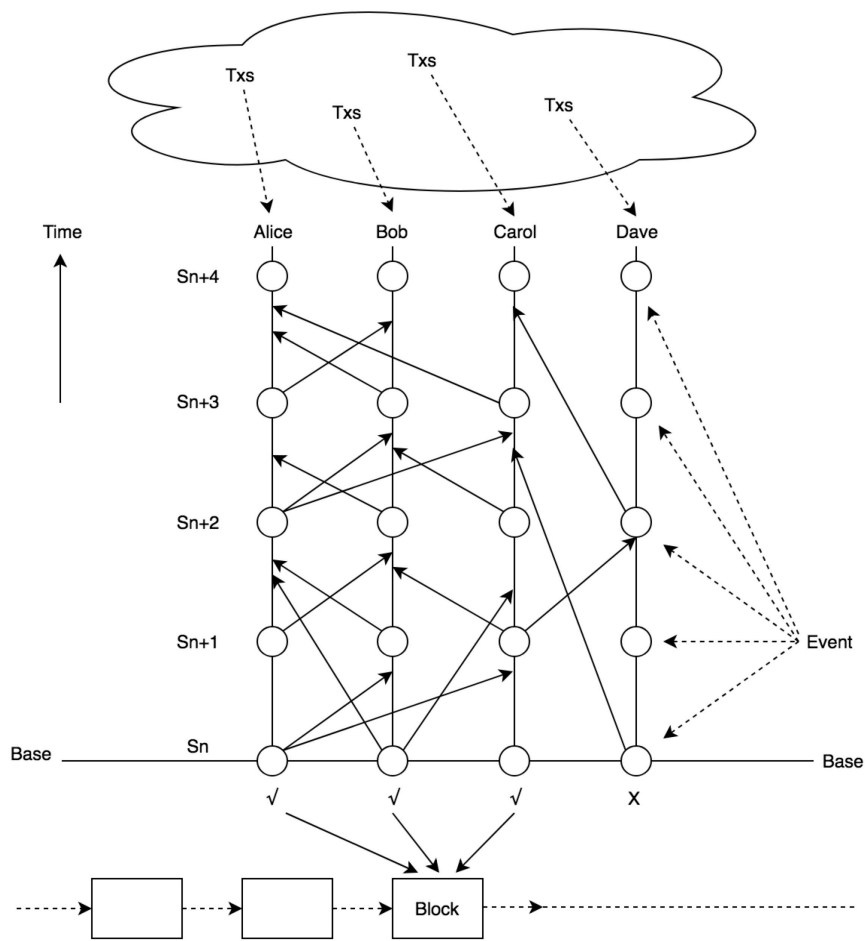

< Figure 5.1 >

In a system with unrestricted number of nodes connected via a p2p network. For a certain period of time, a fixed number of nodes are selected as members of Tetris. We call them **validators**. Validators are special nodes that continuously interact with each other to reach consensus on a set of transactions to be processed and generate new blocks in the system. When a validator join the Tetris, a unique member id was assigned to it called *VID*.

How are the validators be selected, and how an old validator quit and how can a new validator join, will be treated at upper level protocol called *member rotation protocol*. We'll describe it later.

Each validator continuously accept *transactions* and *events* broadcasted on the p2p network from all other nodes on the network. Each *transaction* and *event* can be accepted by a number of validators.

Periodically, the validator create an **event**, and broadcast it to all other validators. An **event** contains all the hashes of transactions received during the period, and all the hashes of its *parents* which defined as:

**Definition 5.1. Parents,** *Parents* of an event comprised of *self-parent* and *other-parents*, *self-parent* is the last event this validator created and *other-parents* are all the events received from other validators during the period.

When an *event* was created by a validator, a sequence number *N* was assigned to it.

$$N = \text{Max}(N \text{ of all the parents}) + 1.$$

Sequence number *N* start from 0 and will keep increasing. If *N* of an event is greater than that of its self-parent's plus 1, then empty placeholder events were created between them to keep the sequence number increasing one by one.

So an event is a collection containing the following entries:

- VID: event's creator
- N: a sequence number start from 0
- Hash of all the transactions received during the period.
- Hash of all the events received from other validator during the period.

which denoted as

$$E = \{VID, N, \{\text{Hash of } E0, E1, ...\}, \{\text{Hash of } tx0, tx1, tx2, ...\}\}$$

Where we specify the first item E0 must be the *self-parent*, if it does not exist, then we specify E0 = null and Hash(E0) = 0.

When the new created event to be broadcast, it is also be signed by the validator, so others received the event can verify it.

When an event is received and verified, it include all the hash of its parent events. The validator check these hashes, if the corresponding parent event of the hash has not arrived, it send a request to ask for the event. In our p2p network design, this is done by send a DHT request. Of course, according to our design, every events and transactions will write to the DHT temporarily while broadcasting. This is our *push/pull model* we'll describe at the next section. An event will be accepted until all its *parent* accepted.

Thus, in each validator, all the events it created or received, by their *parents-children* relationship, form a directed graph like Figure 5.1. This graph is an actual data structure stored in the validator, which we can call it a *tetris* and denoted it as $\mathscr{T}$. Furthermore, we using $\mathscr{T}_{vid,n}$ to denote the sub-tetris under the event $E_{vid,n}$. For example, $\mathscr{T}_{alice,4}$, $\mathscr{T}_{bob,3}$, $\mathscr{T}_{carol,4}$, $\mathscr{T}_{dave,3}$ of Figure5.1 look like:

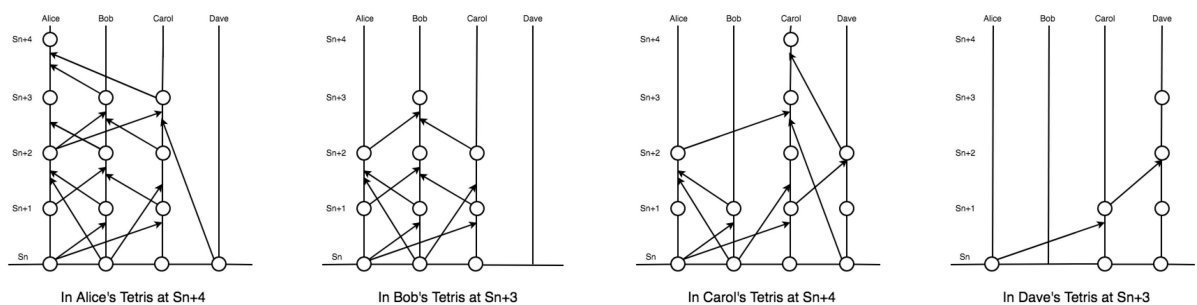

< Figure 5.2 >

What we do here, is first for the purpose to construct a Full-Information Protocol.

**Lemma 5.2.** Tetris is a Full-Information Protocol

Proof: A protocol is said to be a full-information protocol if each process is required to send all its current state to all processes at each step. In Tetris, when a validator broadcasting an event, the event include the hashes of all transactions it received, the event also include hashes of all events it received which also include all the hashes of it's received transactions and predecessor events and so on. Because the hash function is secure, this hash chain can guarantee the whole history is correct when it was received by other validators. other validators using these hashes to retrieve the original information safely, and can know all the state of the sender.

This kind of FIP make validators in Tetris obtain maximum knowledge about the running of the system. For example, in Figure 5.2, Alice at Sn+4 must know that Bob know That Carol know Alice's Sn event. And Carol at Sn+4 must know Alice know that Bob's Sn event. Hashgraph's gossip of gossip[20] can also be seen as a FIP. But their members need directly connect to each other to gossip the information. We establish this FIP on a p2p broadcasting network, and validators even don't know each other, and validators can freely join or quit the system. This is crucial for a permissionless blockchain.

Events and transactions are broadcasted on the network, only their hashes are packed into the body of events to keep the relationship information of them. By this kind of hash chain, Tetris is a FIP with minimum bandwidth requirement.

Any time, the bottom line of events with same sequence number in Figure5.1 are called *base event*, all the *base events* are waiting for a decision for whether it can be committed at this stage. The decision was made based on the continuous coming new events and transactions fall down from upside of the figure. The piled up new events with their relations form some kind of structure that will satisfy certain conditions at a certain level and therefore help making the decision. Until all the *base events* are decided YES/NO to commit, this bottom line is confirmed. All the transactions in the YES decision event will tagged as ready to commit and will later packed into a block. The block's height is same as the sequence number. Then the bottom line of *base events* is eliminated, the above line of events drop down to the bottom line and repeat the above process to produce the next block. We call this process a *stage*, the stage number is equal to the block's height or the sequence number of the *base event*.

This process is very similar to the world famous Russian game Tetris, where new tetriminos fall down continuously and the bottom horizontal line of tetriminos are destroyed automatically when certain conditions are met. This is where the name of our algorithm comes from!

Of course, we are not designing games. What important here is each validator has its own *tetris* $\mathscr{T}$, each validator should independently make a decision base on its own tetris, their decision must be the same, but their $\mathscr{T}$ may not be identical.

At each stage, with the accumulation of knowledge of validators by receive events, validators eventually get enough knowledge to deduce that some form of common knowledge can obtained and then support coordinated decision. What we need to emphasize is that the computing of stages are independent. Because at different stage, old validators might quit, new validators might join. It is another crucial requirement for a permissionless blockchain. For example, at stage n, a consensus decision is made that one validator should quit, then at next stage, all the events from that validator need to be ignored.

The tetris $\mathcal{T}$ of each validator grows up with new events been created and received as time goes on. For the very recent events, at the upper part of the tetris, every validators may have events that the others not yet seen. But for the older events at the bottom part, they will roughly be the same after some time. They may not be identical at particular moment, but they are always consistent. Here consistent means that if there are tetris A of Alice and B of Bob, if A and B both contain event *e*, then they will both contain exactly the same set of ancestors for *e*.

**Definition 5.3. Ancestors,** *Ancestors* of an event *e* include *e* itself and all the *parents* of *e*, all the *parents* of *parents* of *e*, and so on.

**Lemma 5.4.** Tetrises of all validators are consistent

Proof: A and B contain same event *e*, so the hashes of *parents* contain in the *e* must be the same, and all the *parents* must be existed because if *e* to be accepted in a tetris, all his parents according to the hashes must be received and accepted already. And so does the same of *parents* of *parents* and so on. Because the hash function is assumed to be secure, so A and B must contain all the same of ancestors of *e*.

Consistency of tetrises of all validators is the base of our following work when we define functions on $\mathcal{T}$ so same $\mathcal{T}$ at different validators must generate the same output.

Although tetrises are consistent, but according to our assumption of the byzantine system model, some of validators may be byzantine, typically, we assumed that number of faulty nodes does not exceed t, and number of all nodes is 3t+1. A byzantine node may maliciously broadcast forked events, e.g.,create different events and send one to some validators and another to others.

**Definition 5.5. Know,** An event x *know* y, if y is the ancestor of x, and ancestors of x do not include a fork by the creator of y.

**Definition 5.6. Know-well,** An event x *know-well* y, if x *know* y, and there is a set S of events by at least 2t+1 of the members such that x know every event in S, and every event in S know y.

**Lemma 5.7.** At a certain position of tetris, an event e has a fork e', then if there is an event x in one validator tetris *know-well* e, then there are no event in any validators can *know-well* e'

Proof: Suppose tetris A of validator Alice and B of Bob, an event x in A *know-well* e, an event y in B *know-well* e'. The x *know-well* e means there are events of at least 2t+1 members which x *know* all *know* e. So there must exist events from at least t+1 non-faulty members *know* e. In tetris B, event y should *know* events from at least 2t+1 members which *know* e'. Because tetris A and B are consistent, so there must exist at least 1 non-faulty member who has an event *know* both e and e'. According our definition of *know*, this is impossible.

This lemma is similar to the *Strongly Seeing Lemma* in Hashgraph[20] and the *Prepared* status in PBFT[12]. All them doing is just like the *Reliable Broadcast primitive* in Bracha's[13], to restrict the adversary in sending non-fork events, forces him to follow the protocol for events broadcasting.

In each *stage* of the running of Tetris, we'll define the concept of *witness* and *round*.

**Definition 5.8. Round,** At each stage of the running of Tetris, the base events are specified round = 0. For each event e above the base events, We specify its *round* = r + i, where r is the maximum round number of the parents of e, and i = 1 if e can *know-well* at least 2t+1 witnesses in round r, or i = 0 otherwise.

**Definition 5.9 Witness,** Witness is the first event created by a validator in a round. The base events are called round 0 witness.

For example, in Figure 5.1, event $E_{bob,sn+4}$ is a round 1 witness because it know-well $E_{alice,sn}$, $E_{bob,sn}$, and $E_{Carol,sn}$.

Up to here, we have transformed the chaotic relationship of events like shown in Figure 5.1, into a relatively ordered structure as:

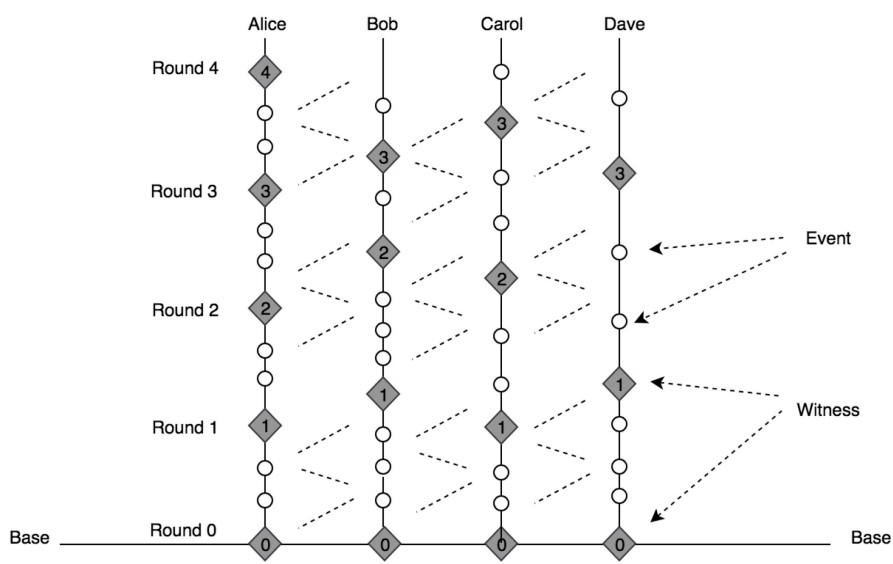

< Figure 5.3 >

This structure to some extent is similar to the round concept of synchronous system, where we can see:

1. At every round r, there must exist at least 2t+1 witnesses suppose a round r+1 witness existed.

2. Each round r witness must *know-well* at least 2t+1 round r-1 witnesses.

3. Witnesses at one same round may not have the same sequence number.

4. Witnesses of the Tetris of all validators are consistent according to the Lemmas above.

Having all above things prepared, we are going to discuss how the validators make decision.

As mentioned in Section 4, having a FIP, what we need to do the next step is to designing a decision function $\mathcal{D}(\mathcal{T})$. With the witness defined, our decision function need only defined on witness. For each base event *e*, we define the function as follow:

```
function decide(T, e)

e.committable = UNDECIDED
for each witness w in round 1
    w.vote = true if w know-well e, false otherwise.
for each witness w in round 2
    s = the set of witnesses in round 1 which w know-well
    w.vote = true if there are at least t/2+1 witnesses in s vote
              true otherwise w.vote = false
for r = 3 to current max round
    for each witness w in round r
        s = the set of witnesses in round r-1 which w know-well
        v = majority vote in s, true for a tie
        n = number of events in s with a vote of v
        if n >= 2t+1
            e.committable = v
            w.vote = v
            return v as decided
        else
            w.vote = v
return UNDECIDED
```

This function run repeatedly whenever a new witness found in the Tetris, and finally all the base events has their *committable* property been decided true or false.

Once all the base events been decided, we'll check all the transactions included in these *committable* = true base events. If the hash of one transaction has appeared in at least t+1 *committable* = true base events or their ancestors from different validators, then the transaction can be tagged as committable. All the committable transactions then validated by all the validators and used to generate a new block. Each validator will create a block header for it and sign by its private key and broadcast it. All the nodes on the network once receive t+1 block header signed by validators can be sure to confirm the new generated block. Then this line of base events drop down and a new line of base events repeat all above to generate the next block of the next stage.

There are still a few questions remain to be clarified.

First, why we need hash of a transaction appeared in t+1 validators for tagged as committable? The problem comes from our push/pull traffic mode. Remember that when the validator receive a new event, included it are hashes of its parents and hashes of txs. If the body of the event/tx of these hashes have not been received by the validator, we should retrieve them from dht. So if a byzantine validator send a fake event or txs without store item on the dht network, the receiver will never got these items. As of events, we specify an event can be accepted only when all of its ancestors have accepted to solve the problem. But as of txs, if the same scheme is adopted, the acceptance of event will delay too much, and it is also not necessary when running the consensus algorithm. So here we need t+1 validators, to ensure that at least 1 honest validator send that hash to guarantee the tx and the hash of it is not faked.

The second question is what if there are less than t+1 base events can be decided *committable* = true? Yes, at **Lemma 5.6.13** below, we'll prove that it must exists at least t+1 base events will be decided *committable* = true under our algorithm.

The third question, though tx is broadcasted on the network, when it reaches every validators, great probability not guaranteed that the tx embodied into the events with same sequence number of all validators. Yes, has mentioned above, we'll check the ancestors of the *committable* = true events. these ancestors have sunk down the base line, but we still need them and they are all consistent for all validators. What matters is we need stipulate a depth of how old ancestors can take part in our checking of tx. For example depth of 10 means we'll check tx in ancestors whose sequence number not less than sequence number of base events minus 10. The unlimited depth bring another question of too much sunk ancestors events need to sync while new validator join the group.

The fourth question, we require all the base events's *committable* status been decided before generate new block. But what if some validators's base events have not received and not contained in one's tetris? This is possible, image if one of the validators stop broadcasting it's events, all the tetris of others will not contain it's base event. At **Lemma 5.6.12**, we'll prove that as long as one base event decided, base events that has not appeared must decide false.

Now, it's time to prove the correctness of decision function defined above.

**Lemma 5.10.** For one base event e, if $\mathcal{D}(\mathcal{T}_{alice})$ output decision value v, it will keep output v, never change that value when new event created, new witness found later.

Proof: It is obvious from the algorithm above.

**Lemma 5.11.** For one base event e, if $\mathcal{D}(\mathcal{T}_{alice})$ output decision value v, i.e., e.committable = v. then an eventual common knowledge $C^{\Diamond}\varphi$ obtained where $\varphi$ means e.committable = v.

Proof: If $\mathcal{D}(\mathcal{T}_{alice})$ output v at round r by witness w, then w must *know-well* at least 2t+1 witnesses at round r-1 whose vote = v. For any witness x at round r of other validators, x also *know-well* at least 2t+1 witnesses at round r-1. The number of these witnesses whose vote $\neq$ v must less than t. Then x.vote must be v. Then at r+1 round, all the witnesses must all receive v and can decide v. So when Alice decided v, Alice know that all others must decided v eventually, and Alice also know when other eventually decided v, they know that all others must decided v eventually, and so on. Therefore, $C^{\Diamond}\varphi$ obtained.

Normally, eventual common knowledge can not guarantee consistency according [7]. We should also ensure that when Alice obtain $C^{\Diamond}\varphi$, there will never exists eventual common knowledge of opposite value. In our case, it is obvious. If Bob obtain $C^{\Diamond}\varphi'$ earlier than Alice where $\varphi'$ means e.committable = !v. Then eventually Alice will decided on !v, this contradict with **Lemma 5.10.** as once decided, the decision value will never change.

A valid decision function should satisfy another requirement of termination, $\mathcal{D}$ must eventually output a value v. Our algorithm above, theoretically can not guarantee termination according the FLP theorem[11]. This can avoid by add a coin round periodically on the algorithm above. We can use a cryptographic pseudorandom number or a shared coin protocol with all the above proof still

correct. For example, modified the algorithm as:

```
    for each witness w in round r
        s = the set of witnesses in round r-1 which w know-well
        v = majority vote in s, true for a tie
        n = number of events in s with a vote of v
        c = a constant of interval of coin round, such as 10.
        if r mod c > 0
            if n >= 2t+1
                e.committable = v
                w.vote = v
                return v as decided
            else
                w.vote = v
        else
            if n >= 2t+1
                w.vote = v
            else
                w.vote = middle bit of w.signature
```

But deeper research reveal that such situation can only happened while adversary maliciously split the vote at each step and the attacker should have the capability of fully control the internet and manipulate every message of honest validators. Traditionally, for the sake of theoretical interest, the proposed adversary models usually assume a strong adversary that completely controls the scheduling of the network and decides which processes receive which messages and in what order. In practice, a real adversary usually does not possess this ability, and if it does, it will probably perform simpler attacks like blocking the communication entirely. Therefore, in practice, the network scheduling can be random enough and lead to a fast termination. Two papers show that this is true and that these algorithms can be practical (Moniz et al., 2006a, 2006b)[14].

So, algorithm above can satisfy all the requirements of consensus mentioned above at Section 4.

We still leave several lemma need to be proved here.

**Lemma 5.12.** Any time as long as one base event be decided, base events that have not appeared must be decided *committable* = false.

Proof: Any time when a base event has been decided, either true or false, the decision must be made by some witness above round 3, then at round 2, at least 2t+1 witnesses must existed. All these witnesses must can not *know* base events that have not appeared. So they must vote false at any tetrises of validators as all tetrises are consistent. Then at next round these base events must be decided *committable* = false.

**Lemma 5.13.** At least t+1 base events will be decided *committable* = true every stage.

Proof: First, we assume a tetris $\mathscr{T}$ with 3t+1 base events and 3t+1 round 1 witness. Then the total number of votes of round 1 witness is at least (3t+1)*(2t+1). We consider an extreme situation for distribution of these votes to base events, when t+1 base events get 3t+1 votes each, and the remain 2t base events still has t*(3t+1) votes, on average 1.5t+0.5 for each, so they can not all below 1.5t+0.5. Then consider any round 2 witness, each of them must vote true to the t+1 base events with 3t+1 votes and then this t+1 base events must be decided *committable* = true. for 2t other base events, 1.5t+0.5 votes is the maximum number they can consume and assure that they don't be guaranteely decided *committable* = true.

Then we to reduce the number of base events, or reduce the number of round 1 witness, either satisfy the condition better than above extreme situation. So at least t+1 base events will be decided *committable* = true every stage.

**Theorem 5.14.** Tetris is a valid asynchronous byzantine fault tolerance consensus algorithm.

Proof: According to the lemmas above, all honest validators will eventually make the same decision of the base events's *committable* property under our byzantine system model hypothesis. And at least t+1 base events will be decided *committable* = true every stage. If the hash of one transaction has appeared in at least t+1 *committable* = true base events or their ancestors from different validators, then the transaction must be valid and can be retrieved from dht. All honest validators then can validate these transactions and produce a valid block. Thus, Tetris is a valid asynchronous byzantine fault tolerance consensus algorithm.

## Member rotation protocol

Member rotation protocol is an upper level protocol to change the validators when Tetris running. It is an indispensable basic requirement of permissionless blockchain.

What important first is Tetris it own has the capability of dynamically remove or add validators one by one at any time. Because Tetris run by stages, each stage the decision functions run independently, so when a new stage start, remove a member of Tetris is just ignore all the events of this member, and add a member into the validators need only the new validator start receive and broadcast events and transactions.

Furthermore, Tetris can review the performance of validators and suggest which validator to remove at particular time.

Having these capability of Tetris, upper lever protocol of member rotation can be designed very flexibly and can easily implemented as a pluggable module. So most of PoX currently existed can adopt to do the work.

# 6. Properties

## Performance

Tetris has been tested to support tens of thousands tps. Furthermore, Tetris's high performance has several its own characteristics:

1. Due to the overlay sub-network design of the underlying p2p network of Tetris, the more nodes in the system, more sub-network could be divided, thus the more tps supported.

2. The greater the volume of transactions per unit time, more frequently the events will created and transmit, then quicker to reach consensus. The consensus finality will be reached faster under the heavy load period.

These characteristics different so much from traditional system such as Bitcoin and Ethereum. There the total throughput of the system is limited to the lowest end of the node participated in the system.

## Scalability

Scalability of a blockchain system is shown in the follow aspects:

1. Maximum number of nodes supported by the system
2. Maximum accounts supported by the system
3. Maximum tps of the system
4. Storage requirement of the system

Blockchain based on Tetris can:

1. Support unlimited nodes
2. Support unlimited accounts
3. Tens of thousands of tps
4. Minimum storage requirement for normal nodes.

## Finality and state

In Tetris, once consensus is reached, it's final. So the history is frozen, the state of the system at this point become immutable. This brings great benefits. For a new added node, it may not necessary to sync all the huge history of the blockchain, but only sync with the latest state, thus save a lot of storage space. Furthermore, this finality has the potential for simplify the future cross-blockchain interaction or sharding scheme.

## Fairness

The meaning of fairness here includes two aspects:

1. How the system order the transactions? If Alice and Bob submit a transaction to the system almost at the same time, who will be the first one?

2. Can anybody in the system manipulate the consensus order of transactions? For example, deliberately delay or ignore some of specific transactions?

The exact transaction order does not matter much for some applications, but can be critically important for others. A real world case is in the field of high frequency trading[24], some guys are willing to pay millions of dollars only to reduce the network latency by microseconds for their transactions reach stock exchanges faster than others.

The fairness has been detailed discussed in[20]. Unlike in a centralized system, when consider a distributed peer-to-peer system, the most tricky thing is the definition of "fair".

Obviously, the fair decision of transaction order can not depend on the timestamp since there are no global time in asynchronous systems and simple timestamps are easily forged.

With the algorithms exist leader, such as PBFT, round-robin system, or Pow system, can "fair" be defined as reflecting the order in which the transactions reached the current leader? Unfortunately, It's still a bad idea. The leader could arbitrarily decide to delay or ignore some of transactions, no single individual can be trusted.

In Tetris, there are two conditions need to be fulfilled for a transaction to be committed. First, transaction should be received by t+1 validators. Then these validators should actively participate in the interaction between other validators and its base event have a chance to be decided as *committable* = true event. Fairness defined here reflects the speed of transactions reach a certain number of validators and the activity and performance of these validators in the group. It sounds better in a distributed environment.

# 7. Practical Issue

### P2P network

Tetris relies heavily on the underlying network. To exert the full potential of Tetris, we have several special requirements for the underlying p2p network in permissionless environment.

1. Overlay sub-network. It's crucial to have the ability to create and manage Overlay sub-network in p2p network. In a flatten p2p network, each nodes could receive all the traffic, it can not support a high throughput system.

2. Erasure code enabled DHT. Huge amounts of data will produce in high throughput system. The storage efficiency and economy arisen as an important problem, and we think this can be tackled by some kind of coding.

3. Temp DHT. In Tetris, transactions and events broadcasting on the network, but also need to retrieve in a certain period of time in our push/pull traffic mode. So this kind of data need temporarily store on the DHT and can be retrieved by who need it later.

We will develop a new p2p network for Tetris, and the topic may be elaborated in a separate whitepaper.

### Sentry Node

In a permissionless blockchain, validators of Tetris should hide themselves on the network to avoid possible attacks like DDOS. Although the validators have no direct network connections to other validators by broadcasting message on a p2p network. The few normal nodes directly connected to the validators in p2p network might have the chance to analyze the traffic mode and locate the ip address of the validators. We can avoid this situation by deploy a set of *Sentry Nodes* which connect to the p2p network on behalf of the validators and validators only connect to their *Sentry Nodes*.

# 8. Conclusions

A new Asynchronous Byzantine Fault Tolerance consensus algorithm, Tetris is proposed here for future blockchain infrastructure with its great performance and proved safety.

Tetris algorithm / protocol is constructed mainly under the most challenging permissionless blockchain context. It is easy to port to permission blockchain or other application areas.

We also provide a demo project at https://github.com/yeeco/tetris_demo implementing the basic functions of Tetris, and may be of interest.

## Reference

[1] R. Fagin, J. Y. Halpern, Y. Moses, and M. Y. Vardi. Reasoning about Knowledge. MIT Press, Cambridge, Mass., 1995

[2] J. Y. Halpern and Y. Moses. Knowledge and Common Knowledge in a Distributed Environment. JACM'90.

[3] Ajay D. Kshemkalyani , Mukesh Singhal, Distributed Computing: Principles, Algorithms, and Systems, Cambridge University Press, New York, NY, 2008

[4] C. Dwork and Y. Moses. Knowledge and common knowledge in a Byzantine environment: crash failures. Information and Computation, 88(2):156-186, 1990.

[5] P. Panangaden and S. Taylor. Concurrent common knowledge: define agreement for asynchronous systems. Distributed Computing, 6(2):73-93, 1992

[6] K. M. Chandy and J. Misra. How processes learn. Distributed Computing, 1(1)40-52, 1986.

[7] Joseph Y. Halpern, Yoram Moses, and Orli Waarts. A characterization of eventual Byzantine agreement. In Proceedings of the Ninth ACM Symposium on Principles of Distributed Computing, pages 333–346, August 1990.

[8] MOSES, Y., AND TUTTLE, M.R. Programming simultaneous actions using common knowledge. Algorithmica 3 (1988), 121-169.

[9] Gil Neiger. Using knowledge to achieve consistent coordination in distributed systems. In preparation, July 1990.

[10] L. Lamport, R. Shostak, and M. Pease, "The byzantine generals problem," ACM Trans. Program. Lang. Syst., vol. 4, pp. 382–401, July 1982.

[11] M. J. Fischer, N. A. Lynch, and M. S. Paterson, "Impossibility of distributed consensus with one faulty process," J. ACM, vol. 32, pp. 374–382, Apr. 1985.

[12] M. Castro and B. Liskov. Practical Byzantine Fault Tolerance, Proceedings of the Third Symposium on Operating Systems Design and Implementation. New Orleans, Louisiana, USA, 1999, pp. 173–186.

[13] G. Bracha, Asynchronous byzantine agreement protocols, Inf. Comput., vol. 75, pp. 130–143, Nov. 1987.

[14] M. Correia, G. S. Veronese, N. F. Neves, and P. Veríssimo, "Byzantine consensus in asynchronous message-passing systems: a survey," IJCCBS, vol. 2, no. 2, pp. 141– 161, 2011.

[15] S. Toueg, Randomized byzantine agreements, in Proceedings of the Third Annual ACM Symposium on Principles of Distributed Computing, PODC 1984, (New York, NY, USA), pp. 163–178, ACM.

[16] A. Clement, E. Wong, L. Alvisi, M. Dahlin, Making Byzantine Fault Tolerant Systems Tolerate Byzantine Faults. NSDI, 09

[17] S. Nakamoto, Bitcoin: A peer-to-peer electronic cash system, 2008. [Online]. Available: http://bitcoin.org/bitcoin.pdf

[18] E. Buchman. Tendermint: Byzantine fault tolerance in the age of blockchains. M.Sc. Thesis, University of Guelph, Canada, June 2016.

[19] S. Popov. The tangle. White paper, available at https://iota.org/IOTA_Whitepaper.pdf, 2016.

[20] Baird L. The Swirlds Hashgraph Consensus Algorithm: Fair, Fast, Byzantine Fault Tolerance, Swirlds Tech Report SWIRLDS-TR-2016-01(2016)

[21] NXT Whitepaper, 2014. [Online]. Available: https://wiki.nxtcrypto.org/ wiki/Whitepaper:Nxt#Proof*of*Stake

[22] BitShares Delegated Proof of Stake, 2014. [Online]. Available: https: //github.com/BitShares/bitshares/wiki/Delegated-Proof-of-Stake

[23] Don Tapscott, Alex Tapscott, Blockchain Revolution: How the Technology Behind Bitcoin Is Changing Money, Business, and the World. Brilliance Audio, 2017

[24] https://www.aldec.com/en/company/blog/169--the-race-to-zero-latency-for-high-frequency-trading

[25] Terence Tao, https://terrytao.wordpress.com/2011/05/19/epistemic-logic-temporal-epistemic-

logic-and-the-blue-eyed-islander-puzzle-lower-bound/